\documentclass[twocolumn]{aastex631}
\usepackage{amsmath}
\usepackage{booktabs}
\usepackage{subfigure}
\usepackage{graphicx}

\usepackage{threeparttable}

\shorttitle{oxygen effect on age}
\shortauthors{Sun et al.}
\graphicspath{{./}{figures/}}

\begin{document}









\title{Age of FGK Dwarfs Observed with LAMOST and GALAH: Considering the Oxygen Enhancement}

\author{Tiancheng Sun} 
\affiliation{Institute for Frontiers in Astronomy and Astrophysics, Beijing Normal University,  Beijing 102206, China}
\affiliation{Department of Astronomy, Beijing Normal University, Beijing 100875, People’s Republic of China}


\author{Zhishuai Ge}
\affiliation{Beijing Planetarium, Beijing Academy of Science and Technology, Beijing, 100044, China}
\email{gezhishuai@bjp.org.cn}

\author{Xunzhou Chen}
\affiliation{Research Center for Intelligent Computing Platforms, Zhejiang Laboratory, Hangzhou 311100, China}
\email{cxz@zhejianglab.com}

\author{Shaolan Bi}
\affiliation{Institute for Frontiers in Astronomy and Astrophysics, Beijing Normal University,  Beijing 102206, China}
\affiliation{Department of Astronomy, Beijing Normal University, Beijing 100875, People’s Republic of China}
\email{bisl@bnu.edu.cn}

\author{Tanda Li}
\affiliation{Institute for Frontiers in Astronomy and Astrophysics, Beijing Normal University,  Beijing 102206, China}
\affiliation{Department of Astronomy, Beijing Normal University, Beijing 100875, People’s Republic of China}
\affiliation{School of Physics and Astronomy, University of Birmingham, Birmingham, B15 2TT, United Kingdom}

\author{Xianfei Zhang}
\affiliation{Institute for Frontiers in Astronomy and Astrophysics, Beijing Normal University,  Beijing 102206, China}
\affiliation{Department of Astronomy, Beijing Normal University, Beijing 100875, People’s Republic of China}

\author{Yaguang Li}
\affiliation{Sydney Institute for Astronomy (SIfA), School of Physics, University of Sydney, NSW 2006, Australia}

\author{Yaqian Wu}
\affiliation{Key Laboratory of Optical Astronomy, National Astronomical Observatories, Chinese Academy of Sciences, A20\\
Datun Rd., Chaoyang District, Beijing 100101, 
People’s Republic of China}

\author{Sarah A. Bird}
\affiliation{Center for Astronomy and Space Sciences, China Three Gorges University, Yichang 443002, People's Republic of China}

\author{Ferguson J. W.}
\affiliation{Department of Physics, Wichita State University, Wichita, KS 67260-0032, USA}

\author{Jianzhao Zhou}
\affiliation{Institute for Frontiers in Astronomy and Astrophysics, Beijing Normal University,  Beijing 102206, China}
\affiliation{Department of Astronomy, Beijing Normal University, Beijing 100875, People’s Republic of China}

\author{Lifei Ye}
\affiliation{Institute for Frontiers in Astronomy and Astrophysics, Beijing Normal University,  Beijing 102206, China}
\affiliation{Department of Astronomy, Beijing Normal University, Beijing 100875, People’s Republic of China}

\author{Liu Long}
\affiliation{Institute for Frontiers in Astronomy and Astrophysics, Beijing Normal University,  Beijing 102206, China}
\affiliation{Department of Astronomy, Beijing Normal University, Beijing 100875, People’s Republic of China}

\author{Jinghua Zhang}
\affiliation{Key Laboratory of Optical Astronomy, National Astronomical Observatories, Chinese Academy of Sciences, A20\\
Datun Rd., Chaoyang District, Beijing 100101, 
People’s Republic of China}



\begin{abstract}
Varying oxygen abundance could impact the modeling-inferred ages. This work aims to estimate the ages of dwarfs considering observed oxygen abundance. To characterize 67,503 LAMOST and 4,006 GALAH FGK-type dwarf stars, we construct a grid of stellar models which take into account oxygen abundance as an independent model input. Compared with ages determined with commonly-used $\alpha$-enhanced models, we find a difference of $\sim$9\% on average when the observed oxygen abundance is considered. The age differences between the two types of models are correlated to [Fe/H] and [O/$\alpha$], and they are relatively significant on stars with [Fe/H] $\lesssim$ $-$0.6 dex. Generally, varying 0.2 dex in [O/$\alpha$] will alter the age estimates of metal-rich ($-$0.2 $<$ [Fe/H] $<$ 0.2) stars by $\sim$10\%, and relatively metal-poor ($-$1 $<$ [Fe/H] $<$ $-$0.2) stars by $\sim$15\%. Of the low-O stars with [Fe/H] $<$ 0.1 dex and [O/$\alpha$] $\sim$ $-$0.2 dex, many have fractional age differences of $\geq$ 10\%, and even reach up to 27\%. The fractional age difference of high-O stars with [O/$\alpha$] $\sim$ 0.4 dex reaches up to $-$33\% to $-$42\% at [Fe/H] $\lesssim$ $-$0.6 dex. We also analyze the chemical properties of these stars. We find a decreasing trend of [Fe/H] with age from 7.5--9 Gyr to 5--6.5 Gyr for the stars from the LAMOST and GALAH. The [O/Fe] of these stars increases with decreasing age from 7.5--9 Gyr to 3--4 Gyr, indicating that the younger population is more O-rich.
\end{abstract}



\section{Introduction} \label{sec:intro}

Galactic archaeology uses the chemical abundances, kinematics, and derived ages of resolved stellar populations as fossils to investigate the formation and evolution history of the Milky Way \citep{2002ARA&A..40..487F, 2020ARA&A..58..205H}. However, in comparison to chemical abundance and kinematics estimation, estimating the ages of field stars is a challenging task due to the inherent uncertainties present in both observational data and the stellar models employed for dating stars \citep{2010ARA&A..48..581S}.

The chemical composition of a star is a fundamental input parameter in the construction of its theoretical model, which is critical in the determination of its age. Notably, at fixed [Fe/H], the abundance variations of individual elements exert a consequential impact on the overall metallicity Z, which subsequently determines the opacity of the stellar models. This, in turn, influences the efficiency of energy transfer and the thermal structure, thereby altering the evolution tracks on the HR diagram and the main-sequence lifetime \citep[e.g.,][]{2012ApJ...755...15V,2022ApJ...929..124C}. Consequently, in the context of stellar modeling, it is essential to consider the proper metal mixture in order to accurately characterize stars and determine their ages.
The solar-scaled ([$\alpha$/Fe] = 0) and $\alpha$-enhanced mixtures have been commonly used in theoretical model grids like Y2 isochrones \citep{2001ApJS..136..417Y,2003ApJS..144..259Y,2002ApJS..143..499K,2004ApJS..155..667D}, Dartmouth Stellar Evolution Database \citep{2008ApJS..178...89D}, and Padova stellar models \citep{2000A&AS..141..371G,2000A&A...361.1023S,2012MNRAS.427..127B,2018MNRAS.476..496F}. These models treated all the $\alpha$-elements, that are O, Ne, Mg, Si, S, Ca, Ti, by the same factor.

Observations from high-resolution spectroscopic data have presented very different O-enhancement values from other $\alpha$-elements on many stars  \citep{2005A&A...433..185B, 2006MNRAS.367.1329R, 2014A&A...568A..25N, 2015A&A...576A..89B,2019A&A...630A.104A}. 
The observed discrepancies in the abundances of oxygen and other $\alpha$-elements can be attributed to the diverse origins of these elements. Specifically, O and Mg are believed to be primarily synthesized during the hydrostatic burning phase of massive stars and subsequently ejected during the core-collapse supernovae (CCSNe) \citep[e.g.,][]{1989ARA&A..27..279W,2006ApJ...653.1145K,2020ApJ...900..179K}. Nevertheless, some works have provided evidence that Mg might also be partially released into the interstellar medium by SNe Ia \citep{2017A&A...603A...2M,2018MNRAS.477.1206N,2018MNRAS.477..438V,2021AJ....161....9F}, while O appears to be solely enriched by CCSNe \citep{2021AJ....161....9F}. The other $\alpha$-elements, namely Si, Ca, and Ti, primarily originate from the explosive burning of CCSNe and are partially contributed by SNe Ia \citep[e.g.,][]{2005ApJ...623..213C,2012MNRAS.426.3282M,2020ApJ...900..179K}. 
For instance, 22\% of Si and 39\% of Ca come from SNe Ia according to the chemical evolution models in \citet{2020ApJ...900..179K}.
Therefore, not all $\alpha$-elements vary in lockstep, the abundance of oxygen may not necessarily correlate with the abundance of other $\alpha$-elements.

Many works have also discussed the effects of varying individual element abundances on the stellar evolution models \citep{2007ApJ...666..403D,2009ApJ...697..275P,2012ApJ...755...15V,2016ApJ...826..155B}. Theoretical models showed that the oxygen abundance influences the stellar evolution differently from the other $\alpha$-elements \citep{1992ApJ...391..685V,2012ApJ...755...15V}. 
Furthermore, \citet{2016ApJ...833..161G} proposed the CO-extreme models, which treat oxygen abundance differently from the other $\alpha$-elements and add carbon abundance in the stellar evolution models. The models have been employed to determine the ages of thousands of metal-poor halo stars, disk stars, and main sequence turn-off stars \citep{2016ApJ...833..161G,2020ApJ...889..157C,2022ApJ...929..124C}. These results showed that increasing oxygen abundance leads to smaller age determination for the stars with [Fe/H] $<$ $-$0.2. For the stars with [Fe/H] $<$ $-$0.2 and [O/$\alpha$] $>$ 0.2 dex, the age difference would be about 1 Gyr. Due to the limited sample sizes of previous studies (\citet{2016ApJ...833..161G}, with 70 stars, and \citet{2020ApJ...889..157C}, with 148 stars) or the restricted range of [Fe/H] values \citep[{[Fe/H]} $\gtrsim$ $-$0.1 dex,][]{2022ApJ...929..124C}, there is a pressing need for a large and self-consistent sample to conduct a quantitative analysis regarding the impact of O-enhancement on age determination. 

Recently, millions of stars' individual element abundances have been measured by spectroscopic surveys like LAMOST \citep{2012RAA....12..735D,2012RAA....12..723Z,2014IAUS..298..310L,2015RAA....15.1095L}, APOGEE \citep{2017AJ....154...94M}, and GALAH \citep{2015MNRAS.449.2604D,2021MNRAS.506..150B}. These large sky surveys provide an excellent opportunity to study the effects of oxygen abundance variations on age determinations across a wide range of stellar parameters. To investigate the systematic effects of O-enhancement on age determination, we study the dwarf stars with available oxygen abundance measurements from LAMOST and GALAH. This paper is organized as follows: Section \ref{sec:data} mentions the data selection; Section \ref{sec:method} describes computations of stellar model grids; Section \ref{sec:oxygen_effect} demonstrates ages differences between the O-enhanced models and $\alpha$-enhanced models; the resulting age-abundance trends are presented in Section \ref{sec:discussion}; and the conclusions of this work are drawn in Section \ref{sec:conclusion}.

\section{Target selection} \label{sec:data}

In this work, we make use of spectroscopic data from LAMOST DR5 Value Added Catalogue \citep{2019ApJS..245...34X} and Third Data Release of GALAH \citep[DR3;][]{2021MNRAS.506..150B},
together with astrometric data from Gaia Data Release 3 \citep{2022arXiv220800211G}. 

\subsection{Spectroscopic Data} \label{subsec:lamost}
LAMOST (the Large Sky Area Multi-Object Fiber Spectroscopic Telescope) DR5 Value Added Catalog \citep{2019ApJS..245...34X} contains more than 6 million stars with atmosphere parameters ($T_{\rm eff}$, $\log g$, $V_{mic}$) and chemical abundances of 16 elements (C, N, O, Na, Mg, Al, Si, Ca, Ti, Cr, Mn, Fe, Co, Ni, Cu, and Ba). Measurements of element abundances are based on the \textit{DD–Payne} tool \citep{2017ApJ...849L...9T,2019ApJS..245...34X}, which is a data-driven method that incorporates constraints from theoretical spectral models. 
It is noteworthy that, as discussed by \citet{2018ApJ...860..159T}, the direct derivation of oxygen abundances from atomic oxygen lines or oxygen-bearing molecular lines in low-resolution  (R $\sim$ 1800) LAMOST spectra is unfeasible. Alternatively, CH and CN molecular lines can be utilized for indirect estimation of oxygen abundances, as their strengths are sensitive to the amount of carbon locked up in CO molecules. As a result, the LAMOST oxygen abundances are only available in the cooler stars (T$_{\rm eff}$ $\lesssim$ 5700 K), where the CH and CN lines have sufficient strength to allow a reasonably precise ($\pm{0.10}$ dex) estimate of [O/Fe] \citep{2019ApJS..245...34X}.
Due to the wide age range and the preservation of initial chemical abundances, the main-sequence star could be a good tracer of stellar populations. Therefore, we select the main-sequence stars with available measurements for [Fe/H], [$\alpha$/Fe], and [O/Fe] from the catalog. Firstly, we use some recommended labels ($T_{\rm eff}$$\_$flag = 1, $\log g$$\_$flag = 1, [Fe/H]$\_$flag = 1, ${\rm [X/Fe]}\_$flag\footnote{${\rm [X/Fe]}\_$flag = 1 for 14 elements (C, N, O, Na, Mg, Al, Si, Ca, Ti, Cr, Mn, Fe, Co, Ni).} = 1, qflag$\_$chi2 = good) to select stars with reliable measurements. Afterward, we remove stars with T$_{\rm eff}$ smaller than 5000 K or signal-to-noise ratio (S/N) less than 50 because their [O/Fe] determinations are not robust. \citet{2019ApJS..245...34X} also provided a tag named “qflag$\_$singlestar” to infer whether a star is single or belongs to a binary system. The tag is determined by the deviation significance of the spectroscopic parallax from the Gaia astrometric parallax. When the deviation is less than 3$\sigma$, it suggests an object is a single star. We use this tag to remove all candidate binaries from our sample.
Finally, we choose stars with $\log g>$ 4.1. We lastly select a total of 187,455 unique stars. 

\begin{figure*}[ht!]
\includegraphics[scale=0.7]{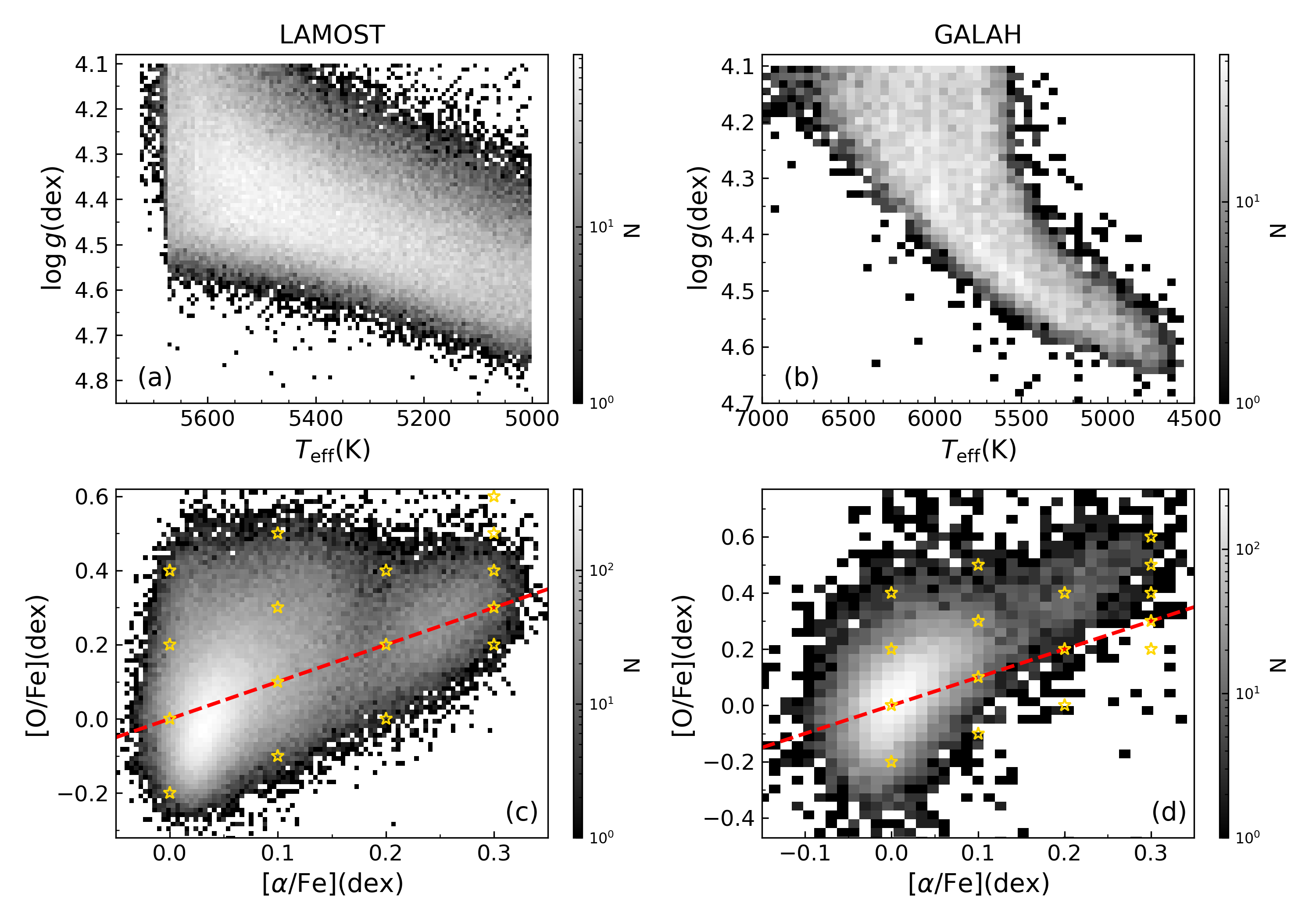}
\caption{Color-coded stellar number density distributions of the targets from LAMOST (left column) and GALAH (right column) in the Kiel diagram (a-b), and the [$\alpha$/Fe]-[O/Fe] space (c-d). The red dashed lines in panels (c)-(d) indicate the 1:1 relation. The gold pentagrams represent the input metal mixtures (shown in Table~\ref{tbl:table1}) of the stellar model grid.    \label{fig:RZ}}
\end{figure*}


GALAH (Galactic Archaeology with HERMES) DR3 \citep{2021MNRAS.506..150B} presents stellar parameters ($T_{\rm eff}$, $\log g$, [Fe/H], $V_{mic}$, $V_{broad}$, $V_{rad}$) and up to 30 elemental abundances for 588,571 stars, derived from optical spectra at a typical resolution of R $\sim$ 28,000.
The oxygen abundance from GALAH DR3 was calculated using the O$_{\rm I}$ 777 nm triplet \citep{2018A&A...616A..89A}, based on a non-LTE method (LTE: local thermodynamic equilibrium)\citep{2020A&A...642A..62A}.
This NLTE method has also been employed for the measurement of [Fe/H] in GALAH.
Following the recommendations in GALAH DR3, we require a SNR $>$ 30, and a quality flag = 0 for reliable stellar parameter determination including iron, $\alpha$-elements, and oxygen abundances (flag$\_$sp = 0, flag$\_$fe$\_$h = 0, flag$\_$alpha$\_$fe = 0, and flag$\_$o$\_$fe = 0). Additionally, the sample is limited to the stars with e$\_$alpha$\_$fe $<$ 0.1 and e$\_$o$\_$fe $<$ 0.1. We exclude the binary systems identified by \citet{2020A&A...638A.145T} (which is a catalog of FGK binary stars in GALAH). These cuts give us a sample of 19,512 dwarf stars ($\log g>$ 4.1).

\subsection{Astrometric Data} \label{subsec:gaia}

We cross-match our selected LAMOST and GALAH samples with Gaia DR3 \citep{2022arXiv220800211G} catalog to obtain the luminosity for each star. Given that luminosity is utilized as a key observational constraint for estimating stellar age, we select stars with luminosity uncertainty less than 10$\%$. Additionally, we select single stars by making a cut based on the Gaia re-normalized unit weight error (RUWE) being less than 1.2 (RUWE values are from the Gaia DR3). Our final sample consists of 149,906 stars from LAMOST (5000 K $<$ $T_{\rm eff}$ $<$ 5725 K, $-$1 $<$ [Fe/H] $<$ 0.5, $\log g>$ 4.1) and 15,591 stars from GALAH (4500 K $<$ $T_{\rm eff}$ $<$ 7000 K, $-$1 $<$ [Fe/H] $<$ 0.5, $\log g>$ 4.1).

We calculate the Galactic Cartesian coordinates (X, Y, Z) and velocities (U, V, W) for the LAMOST sample using the Python package \textit{Galpy} \citep{2015ApJS..216...29B}. The distances are estimated by \citet{2021AJ....161..147B}. The Sun is located at (X, Y, Z) = ($-$8.3, 0, 0) kpc, and the solar motion with respect to the local standard of rest is (U$_{\odot}$, V$_{\odot}$, W$_{\odot}$) = (11.1, 12.24, 7.25) km s$^{-1}$ \citep{2010MNRAS.403.1829S}. We use the Galactic Cartesian coordinates and velocities from the GALAH DR3 value-added catalog (VAC), which is based on astrometry from Gaia EDR3 and radial velocities determined from the GALAH spectra \citep{2021MNRAS.508.4202Z}.

In Figure \ref{fig:RZ}, we demonstrate dwarfs from LAMOST and GALAH in the Kiel diagram, and the [$\alpha$/Fe]\footnote{The [$\alpha$/Fe] from both the LAMOST and GALAH catalog are defined as an error-weighted mean of [Mg/Fe], [Si/Fe], [Ca/Fe] and [Ti/Fe].}-[O/Fe] space to inspect their general distributions.
The Kiel diagram in Figure \ref{fig:RZ}(a) shows that most of the LAMOST dwarfs are cooler than 5700 K, while the GALAH dwarfs in Figure \ref{fig:RZ}(b) covers a wider range of T$_{\rm eff}$ (4500 - 7000 K). It should be noted that we do not apply any cut-off value at the high temperature side for the LAMOST sample. This upper limit is where reliable oxygen abundance can be measured by \citet{2019ApJS..245...34X}. 
The [$\alpha$/Fe]-[O/Fe] diagrams in Figure \ref{fig:RZ}(c-d) show that the [O/Fe] generally increases with increasing [$\alpha$/Fe], however, [O/Fe] widely spread at given $\alpha$-enhanced values. The spreading is relatively large for low-$\alpha$ stars (especially for the GALAH sample), ranging from $-$0.4 to +0.6.

\begin{deluxetable}{c r c c}[ht!]
\label{tbl:table0}
\tablecaption{Metal Mixtures for the GS98 Solar Mixture, the $\alpha$-Enhanced Mixture, and the O-Enhanced Mixture.}
\setlength{\tabcolsep}{10pt}
\tabletypesize{\small}
\tablehead{
\colhead{Element} & \colhead{$\log N_{\odot}$} & \colhead{$\log N_{\alpha \rm EM}$} & \colhead{$\log N_{\rm OEM}$} }
\startdata
C & 8.52 & 8.52 & 8.52 \\
N & 7.92 & 7.92 & 7.92 \\
O & 8.83 & 8.83+[$\alpha$/Fe] & 8.83+[O/Fe] \\
F & 4.56 & 4.56 & 4.56 \\
Ne & 8.08 & 8.08+[$\alpha$/Fe] & 8.08+[$\alpha$/Fe] \\
Na & 6.33 & 6.33 & 6.33 \\
Mg & 7.58 & 7.58+[$\alpha$/Fe] & 7.58+[$\alpha$/Fe] \\
Al & 6.47 & 6.47 & 6.47 \\
Si & 7.55 & 7.55+[$\alpha$/Fe] & 7.55+[$\alpha$/Fe] \\
P & 5.45 & 5.45 & 5.45 \\
S & 7.33 & 7.33+[$\alpha$/Fe] & 7.33+[$\alpha$/Fe] \\
Cl & 5.50 & 5.50 & 5.50 \\
Ar & 6.40 & 6.40 & 6.40 \\
K & 5.12 & 5.12 & 5.12 \\
Ca & 6.36 & 6.36+[$\alpha$/Fe] & 6.36+[$\alpha$/Fe] \\
Sc & 3.17 & 3.17 & 3.17 \\
Ti & 5.02 & 5.02+[$\alpha$/Fe] & 5.02+[$\alpha$/Fe] \\
V & 4.00 & 4.00 & 4.00 \\
Cr & 5.67 & 5.67 & 5.67 \\
Mn & 5.39 & 5.39 & 5.39 \\
Fe & 7.50 & 7.50 & 7.50 \\
Co & 4.92 & 4.92 & 4.92 \\
Ni & 6.25 & 6.25 & 6.25 \\
\enddata
\end{deluxetable}

\begin{deluxetable}{c r c}[ht!]
\label{tbl:table1}
\tablecaption{Grid of Evolutionary Models with Two Metal Mixture Patterns.}
\setlength{\tabcolsep}{10pt}
\tabletypesize{\small}
\tablehead{
\colhead{Metal-mixture} & \colhead{[O/Fe]} & \colhead{[$\alpha$/Fe]} \\ 
\colhead{} & \colhead{(dex)} & \colhead{(dex)} }
\startdata
O-enhanced mixture & $-0.2$ & $0$ \\
$ $ & $0.2$ & $0$ \\
$ $ & $0.4$ & $0$ \\
$ $ & $-0.1$ & $0.1$ \\
$ $ & $0.3$ & $0.1$ \\
$ $ & $0.5$ & $0.1$ \\
$ $ & $0$ & $0.2$ \\
$ $ & $0.4$ & $0.2$ \\
$ $ & $0.2$ & $0.3$ \\
$ $ & $0.4$ & $0.3$  \\
$ $ & $0.5$ & $0.3$  \\
$ $ & $0.6$ & $0.3$  \\
\hline 
$\alpha$-enhanced mixture & $0$ & $0$ \\
$ $ & $0.1$ & $0.1$  \\
$ $ & $0.2$ & $0.2$ \\
$ $ & $0.3$ & $0.3$  \\
\enddata
\end{deluxetable}

\begin{deluxetable}{r c c c}[ht!]
\label{tbl:table2}
\tablecaption{Z Values of Fixed [Fe/H] with Two Metal Mixture Patterns.}
\setlength{\tabcolsep}{14pt}
\tabletypesize{\small}
\tablehead{
\colhead{[Fe/H]}  & \colhead{[$\alpha$/Fe]} & \colhead{[O/Fe]} & \colhead{Z} \\ 
\colhead{(dex)} & \colhead{(dex)} & \colhead{(dex)} & \colhead{(dex)} }
\startdata
$-1.0$ & $0.1$ & $0.1$ & $0.0020$\\
$-1.0$ & $0.1$ & $0.5$ & $0.0036$ \\
\hline
$-0.8$ & $0.1$ & $0.1$ & $0.0032$  \\
$-0.8$ & $0.1$ & $0.5$ & $0.0056$ \\
\hline
$-0.6$ & $0.1$ & $0.1$ & $0.0051$ \\
$-0.6$ & $0.1$ & $0.5$ & $0.0089$ \\
\hline
$-0.4$ & $0.1$ & $0.1$ & $0.0080$  \\
$-0.4$ & $0.1$ & $0.5$ & $0.0139$ \\
\hline
$-0.2$ & $0.1$ & $0.1$ & $0.0126$ \\
$-0.2$ & $0.1$ & $0.5$ & $0.0217$ \\
\hline
$0$ & $0.1$ & $0.1$ & $0.0197$  \\
$0$ & $0.1$ & $0.5$ & $0.0337$ \\
\enddata
\end{deluxetable}

\begin{figure*}[ht!]
\includegraphics[scale=0.49]{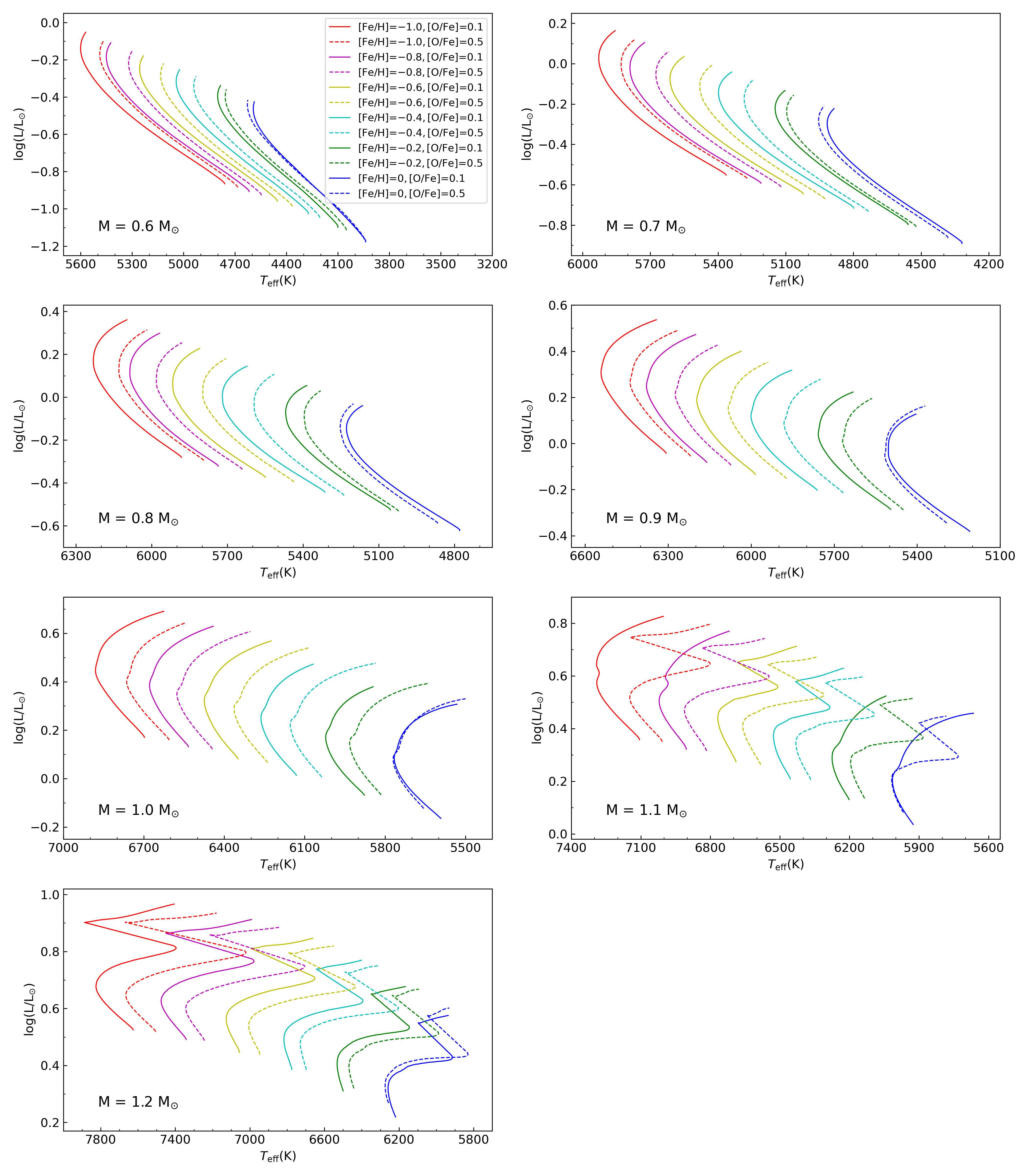}
\caption{Stellar evolution tracks of fixed mass (M = 0.6, 0.7, 0.8, 0.9, 1.0, 1.1, 1.2 M$_{\odot}$) computed with $\alpha$EM and OEM models. In each panel, the [Fe/H] range is 0, $-$0.2, $-$0.4, $-$0.6, $-$0.8, and $-$1.0 (from right to left). The solid lines and dashed lines represent the tracks with input [O/Fe] = 0.1 and 0.5, respectively. All tracks have the same input [$\alpha$/Fe] (0.1 dex) values. The Z values of each track are presented in Table~\ref{tbl:table2}. \label{fig:grid}}
\end{figure*}

\begin{figure*}[ht!]
\includegraphics[scale=0.55]{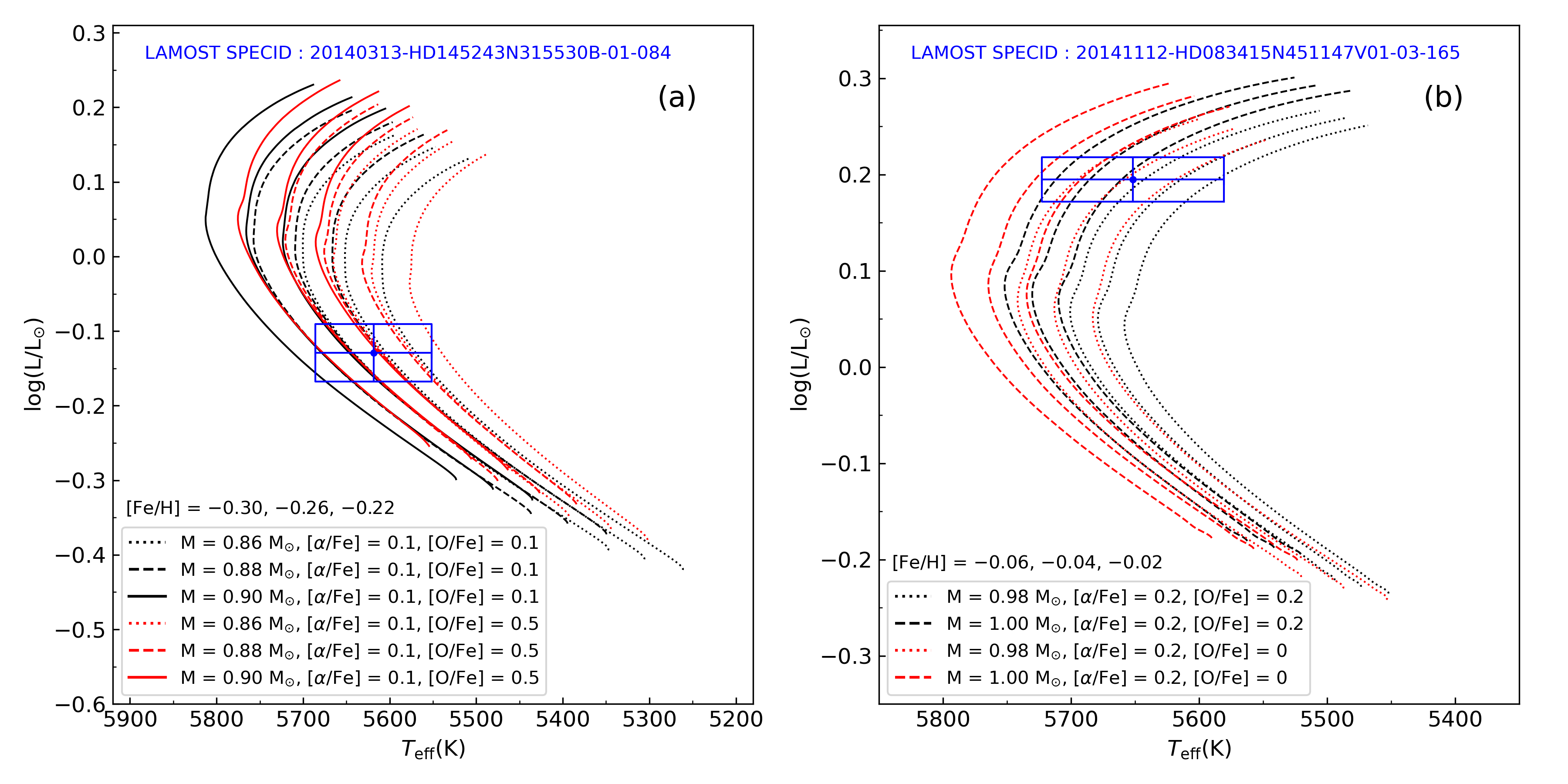}
\caption{Stellar evolution tracks of two example stars calculated with $\alpha$EM (black) and OEM (red) models. Panel (a) shows 
the tracks of a star with [$\alpha$/Fe] $\sim$0.1 and [O/Fe] $\sim$0.5 (LAMOST SPECID: 20140313-HD145243N315530B-01-084); panel (b) shows the tracks of a star with [$\alpha$/Fe] $\sim$0.2 and [O/Fe] $\sim$0 (LAMOST SPECID: 20141112-HD083415N451147V01-03-165). At fixed mass and metal mixture (fixed color and line type), the input [Fe/H] values of tracks (from left to right) are shown. 
The blue dots and error bars are the observed values and uncertainties of example stars, while the blue squares represent the observational error box. The atmosphere parameters and chemical abundance for these stars are shown in table \ref{tbl:table_lamost}.
\label{fig:grid-fit}}
\end{figure*}

\begin{deluxetable*}{c c c c c c}[ht!]
\label{tbl:table_lamost}
\tablecaption{Atmosphere Parameters  and Chemical Abundance for the Example Stars from LAMOST}
\setlength{\tabcolsep}{4pt}
\tabletypesize{\small}
\tablehead{
\colhead{Star}  & \colhead{$T_{\rm eff}$} & \colhead{[Fe/H]} & \colhead{Luminosity} & \colhead{[$\alpha$/Fe]} & \colhead{[O/Fe]} 
\\ 
\colhead{sobject$\_$id} & \colhead{(K)} & \colhead{(dex)} & \colhead{(L$_{\odot}$)} & \colhead{(dex)} & \colhead{(dex)} }
\startdata
20140313-HD145243N315530B-01-084 & $5619\pm{22}$ & $-0.30\pm{0.04}$ & $0.74\pm{0.02}$ & $0.06\pm{0.02}$ & $0.46\pm{0.09}$  \\
20141112-HD083415N451147V01-03-165 & $5652\pm{24}$ & $-0.15\pm{0.04}$ & $1.57\pm{0.03}$ & $0.15\pm{0.02}$ & $-0.02\pm{0.08}$ \\
\enddata
\end{deluxetable*}

\begin{figure*}[ht!]
\centering
\subfigure[GALAH sobject$\_$id: 171230005802396]{
\includegraphics[scale=0.6]{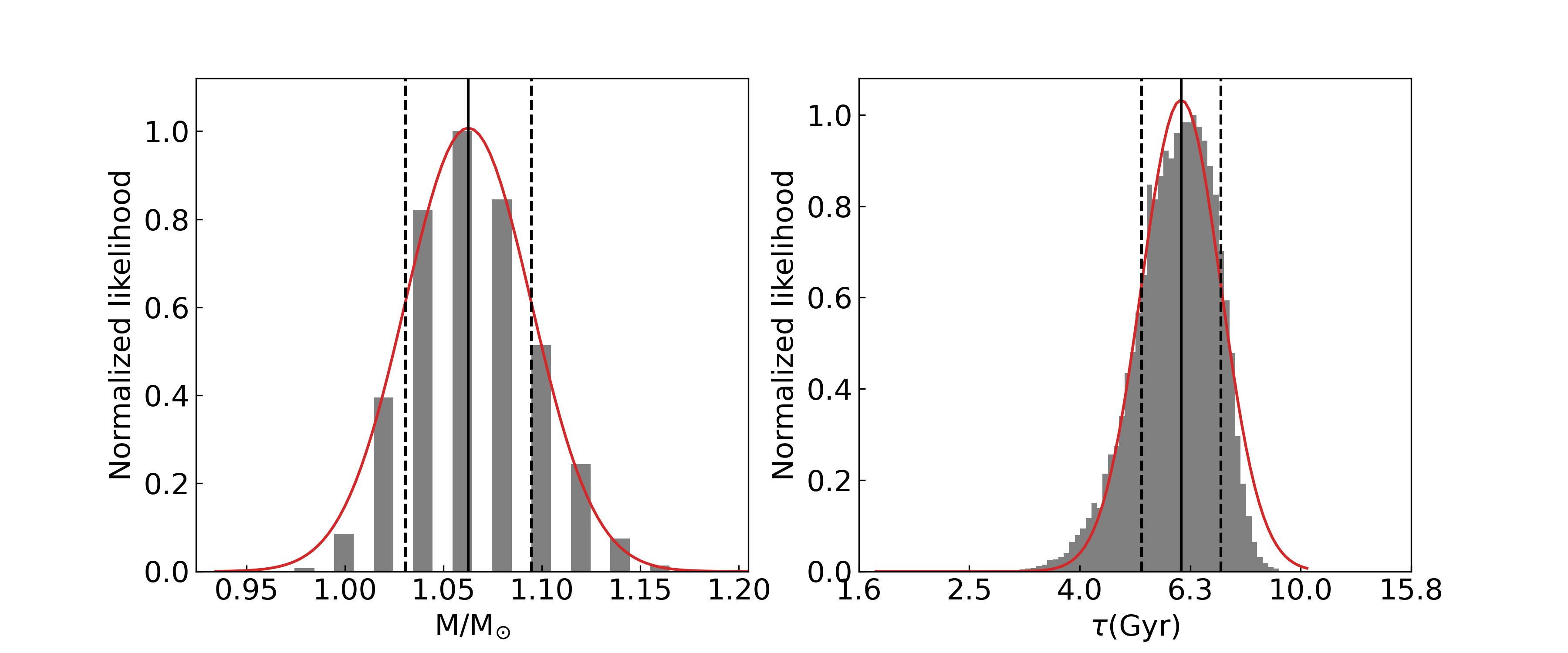}
}
\vskip -6pt
\subfigure[GALAH sobject$\_$id: 160529003401378]{
\includegraphics[scale=0.6]{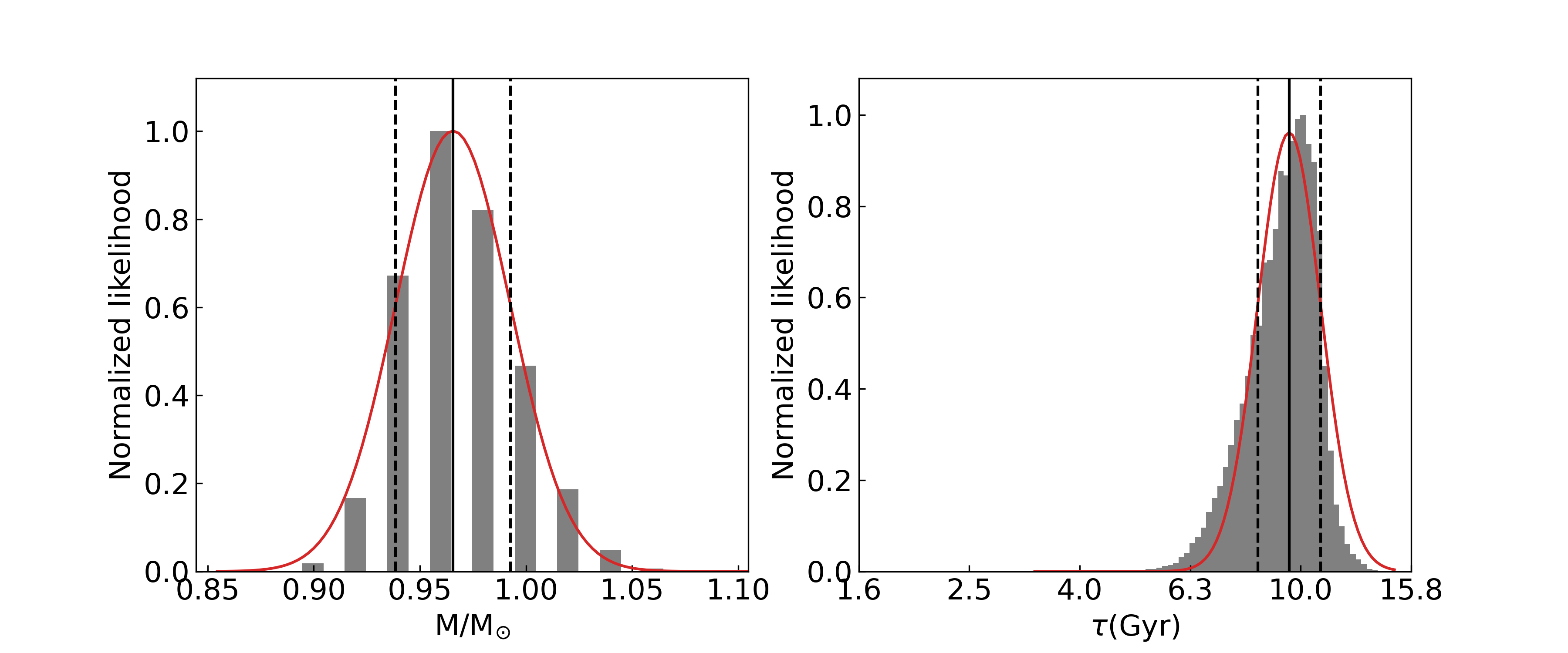}
}
\caption{Likelihood distributions of mass and age for two example stars from GALAH. The red solid line represents the Gaussian function that fits to the likelihood distribution. Solid and dashed vertical lines indicate the mean and standard deviation of the Gaussian profile. The fundamental parameters and chemical abundance for these stars are shown in table \ref{tbl:table3}.
\label{fig:age-fit}}
\end{figure*}

\section{Stellar Models} \label{sec:method}

\subsection{Input Physics} \label{subsec:Input physics}

We construct a stellar model grid using the Modules for Experiments in Stellar Astrophysics (MESA) code \citep [MESA][]{2011ApJS..192....3P, 2013ApJS..208....4P, 2015ApJS..220...15P, 2018ApJS..234...34P, 2019ApJS..243...10P}. The versions of MESA and MESA SDK we used are Revision 12115 and Version 20.3.1, respectively. 

The EOS (Equation of State) tables in MESA are a blend of OPAL \citep {2002ApJ...576.1064R}, SCVH \citep{1995ApJS...99..713S}, PTEH \citep{1995MNRAS.274..964P}, HELM \citep{2000ApJS..126..501T}, and PC \citep{2010CoPP...50...82P} EOS tables. Nuclear reaction rates are a combination of rates from NACRE \citep {1999NuPhA.656....3A}, JINA REACLIB \citep {2010ApJS..189..240C}, plus additional tabulated weak reaction rates \citep{1985ApJ...293....1F, 1994ADNDT..56..231O, 2000NuPhA.673..481L}. Screening is included via the prescription of \citet {2007PhRvD..76b5028C}. Thermal neutrino loss rates are from \citet {1996ApJS..102..411I}. The helium enrichment law is calibrated with initial abundances of helium and heavy elements of the solar model given by \citet{2011ApJS..192....3P}, and it results in $Y$ = 0.248 + 1.3324 $Z$. The mixing-length parameter $\alpha_{\rm MLT}$ is fixed to 1.82. Microscopic diffusion and gravitational settling of elements are necessary for stellar models of low-mass stars, which will lead to a modification to the surface abundances and main-sequence (MS) lifetimes \citep[e.g.,][]{2001ApJ...562..521C,2012MNRAS.427..127B}. Therefore, we include diffusion and gravitational settling using the formulation of \cite{1994ApJ...421..828T}. We use the solar mixture GS98 from \citet{1998SSRv...85..161G}. The opacity tables are OPAL high-temperature opacities \footnote{\url{http://opalopacity.llnl.gov/new.html}} supplemented by the low-temperature opacities \citep{2005ApJ...623..585F}. 

We customize metal mixtures by introducing two enhancement factors, one for oxygen and one for all other $\alpha$-elements (i.e., Ne, Mg, Si, S, Ca, and Ti). The two factors are applied in the same way as \citet{2015MNRAS.447..680G} to vary the volume density of element ($\log N$) based on the GS98 solar mixture as presented in Table \ref{tbl:table0}. 
We make a number of opacity tables by varying two enhancement factors according to the ranges of [$\alpha$/Fe] and [O/Fe] values of the star sample. The enhancement values are shown in Table \ref{tbl:table1}. For the mixtures with the same oxygen and $\alpha$-elements enhancement factors, we refer to them as $\alpha$-enhanced mixture ($\alpha$EM), otherwise, as O-enhanced mixture (OEM).

\begin{deluxetable*}{c c c c c c c c c c}[ht!]
\label{tbl:table3}
\tablecaption{Fundamental Parameters and Chemical Abundance for the Example Stars from GALAH}
\setlength{\tabcolsep}{4pt}
\tabletypesize{\small}
\tablehead{
\colhead{Star}  & \colhead{$T_{\rm eff}$} & \colhead{[Fe/H]} & \colhead{Luminosity} & \colhead{[$\alpha$/Fe]} & \colhead{[O/Fe]} & \colhead{Mass$_{\alpha \rm EM}$} & \colhead{Mass$_{\rm Buder2021}$} & \colhead{Age$_{\alpha \rm EM}$} & \colhead{Age$_{\rm Buder2021}$}
\\ 
\colhead{sobject$\_$id} & \colhead{(K)} & \colhead{(dex)} & \colhead{(L$_{\odot}$)} & \colhead{(dex)} & \colhead{(dex)} & \colhead{(M$_{\odot}$)} & \colhead{(M$_{\odot}$)} & \colhead{(Gyr)} & \colhead{(Gyr)} }
\startdata
$171230005802396$ & $6096\pm{76}$ & $-0.23\pm{0.06}$ & $2.26\pm{0.07}$ & $0\pm{0.02}$ & $0.02\pm{0.08}$ & $1.06\pm{0.03}$ & $1.03\pm{0.04}$ & $6.08\pm{1.01}$ & $6.46\pm{1.17}$ \\
$160529003401378$ & $5846\pm{76}$ & $-0.42\pm{0.06}$ & $1.67\pm{0.03}$ & $0.31\pm{0.03}$ & $0.34\pm{0.09}$ & $0.97\pm{0.03}$ & $0.96\pm{0.03}$ & $9.53\pm{1.26}$ & $10.04\pm{1.39}$ \\
\enddata
\begin{threeparttable}
\begin{tablenotes}
\footnotesize
\item[*] The masses (Mass$_{\rm Buder2021}$) and ages (Age$_{\rm Buder2021}$) of the two example stars from the GALAH value-added catalog \citep{2021MNRAS.506..150B} are calculated based on PARSEC stellar isochrones (the PAdova and TRieste Stellar Evolution Code) \citep{2017ApJ...835...77M}.
\end{tablenotes}
\end{threeparttable}
\end{deluxetable*}

\begin{figure*}[ht]
\includegraphics[scale=0.6]{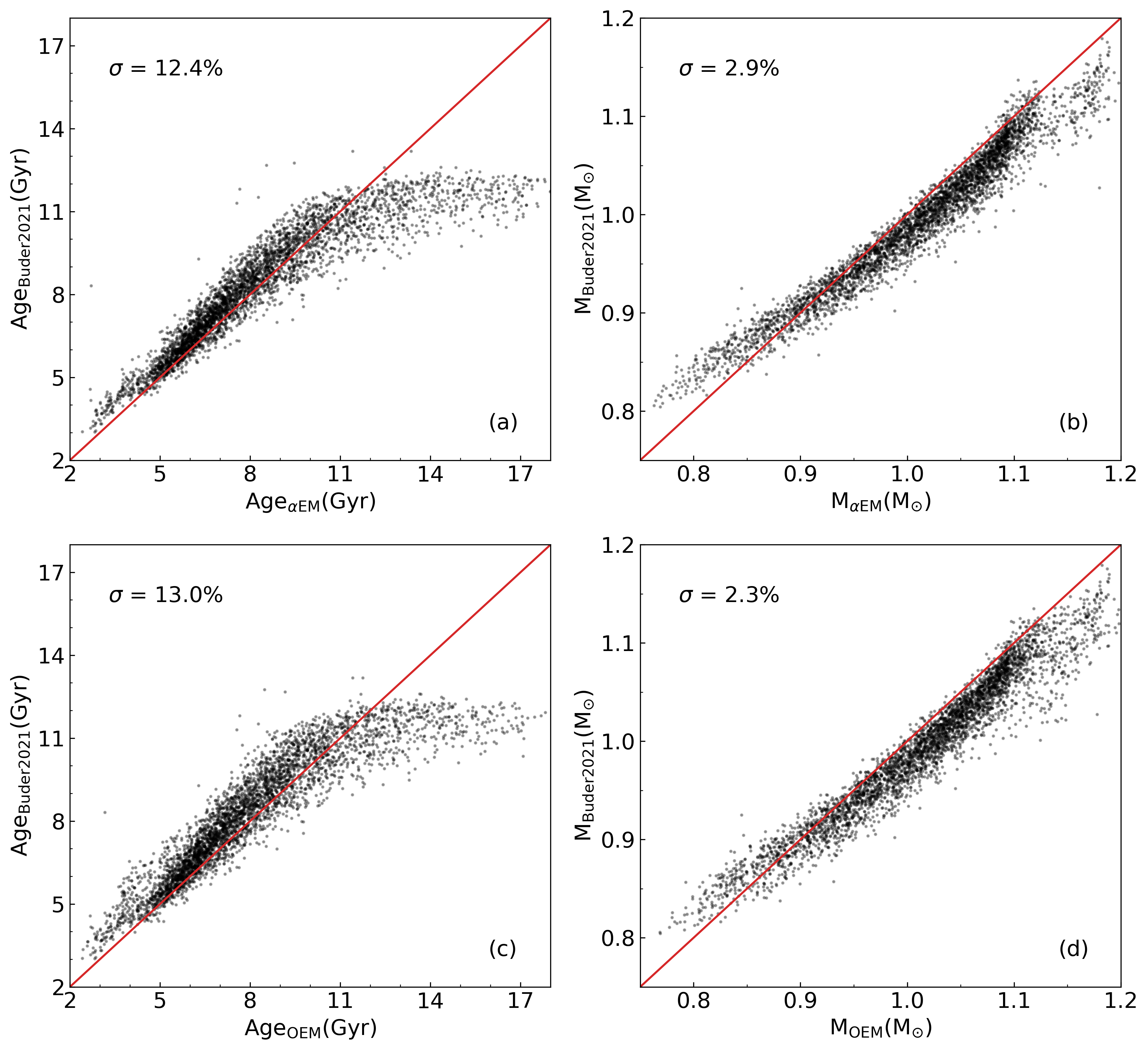}
\caption{Comparison of masses and ages of $\sim$4,000 GALAH sample stars from our $\alpha$EM models, OEM models, and the GALAH DR3 value-added catalog \citep[VAC,][]{2021MNRAS.506..150B}. The red line represents the 1:1 line. Dispersion of the relative age and mass difference is marked in the figure. \label{fig:age_a2}}
\end{figure*}

\begin{figure*}[ht!]
\includegraphics[scale=0.55]{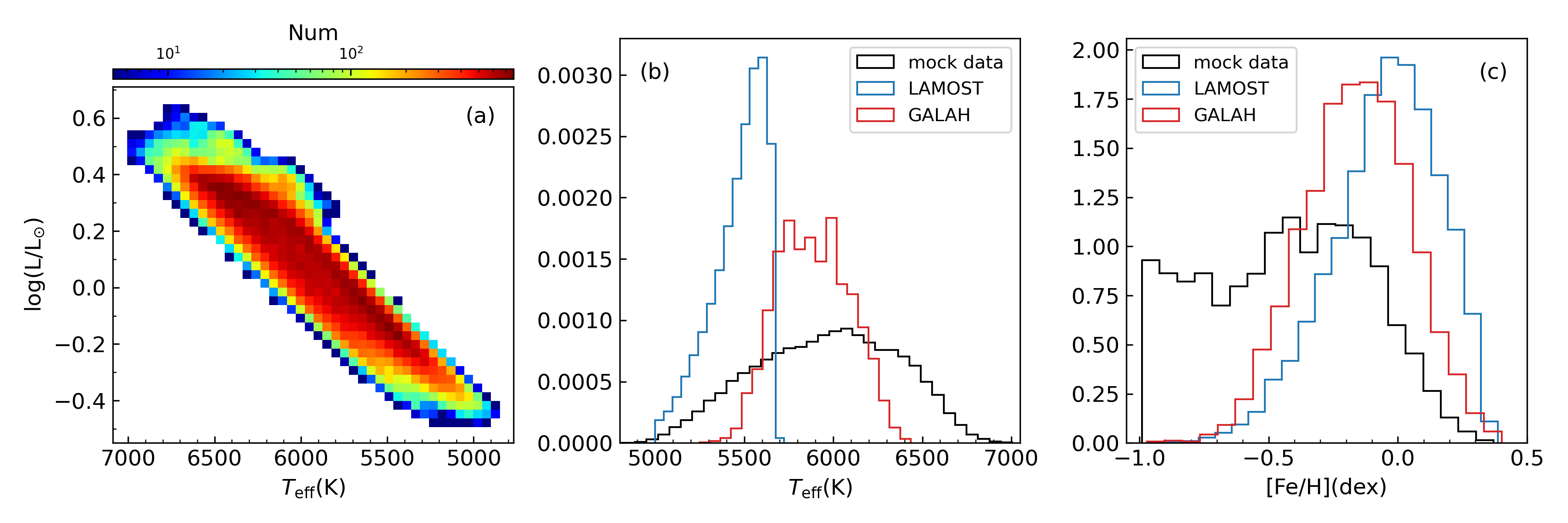}
\caption{Panel (a) shows the color-coded stellar number density distributions of mock stars on the H-R diagram. Panel (b-c) shows the comparison of $T_{\rm eff}$ and [Fe/H] distributions between mock data and observational data. \label{fig:mock_data_test}}
\end{figure*}

\subsection{Grid Computations} \label{subsec:grid}

We establish stellar model grids that include various metal-mixture patterns as indicated in Table~\ref{tbl:table1}. The mass range is from 0.6 to 1.2 M$_{\odot}$ with a grid step of 0.02 M$_{\odot}$. Input [Fe/H] values range from $-$1.20 to +0.46 dex with a grid step of 0.02 dex. The computation starts at the Hayashi line and terminates at the end of main-sequence when core Hydrogen exhausts (mass fraction of center hydrogen goes below 10$^{-12}$).
The inlist file (for MESA) utilized in the computation of our stellar models is available on Zenodo: \dataset[doi:10.5281/zenodo.7866625]{https://doi.org/10.5281/zenodo.7866625}

To explicate the effect of oxygen enhancement on the evolutionary tracks, we provide an exposition of representative evolutionary tracks in Figure \ref{fig:grid}. The corresponding values of Z are listed in Table~\ref{tbl:table2}. At fixed [Fe/H], the variation of [O/Fe] would influence opacity, which could influence the energy transfer efficiency and the thermal structure. 
We find that the larger [O/Fe] leads to higher opacity at input [Fe/H] $\leq$ $-$0.2, and shifts the evolutionary tracks to lower $T_{\rm eff}$.
As seen in Figure \ref{fig:grid}, at [Fe/H] $\leq$ $-$0.2, O-rich models are generally cooler than the $\alpha$-enhanced models at given input [Fe/H], leading to higher modeling-determined masses (smaller ages) for a given position on the HR diagram (left panel of Figure \ref{fig:grid-fit}). However, at input [Fe/H] = 0, 
larger [O/Fe] leads to lower opacity, and shifts the evolutionary tracks to higher $T_{\rm eff}$.
The O-rich models are slightly hotter than the $\alpha$-enhanced models. 
Overall, at fixed mass, the $T_{\rm eff}$ difference between the two models becomes significant with smaller [Fe/H].
In addition, we note that the 1.1 M$_{\odot}$ and 1.2 M$_{\odot}$ tracks of O-rich models show different behavior compared with the tracks of 0.7 $\sim$ 1.0 M$_{\odot}$. The O-rich models with 1.1 M$_{\odot}$ show a blue hook morphology at [Fe/H] $\leq$ $-$0.8, which enlarges the $T_{\rm eff}$ difference between two models at this evolutionary phase. At 1.2 M$_{\odot}$, both models show a blue hook morphology at the end of main-sequence, and the $T_{\rm eff}$ difference keeps approximately constant at [Fe/H] $\leq$ $-$0.6.

Figure \ref{fig:grid-fit} presents the stellar evolution tracks of two example stars calculated with $\alpha$EM and OEM models. Figure \ref{fig:grid-fit}(a) presents the tracks of a star with observed [$\alpha$/Fe] $\sim$ 0.1, [O/Fe] $\sim$ 0.5. Based on the $\alpha$EM models (input [$\alpha$/Fe] = 0.1, [O/Fe] = 0.1), we obtain the best-fit values of fundamental parameters for this star: mass = 0.87 $\pm$ 0.02 M$_{\odot}$, age = 8.69 $\pm$ 1.49 Gyr (the fitting method is described in detail in Section \ref{subsec:Parameter Estimation}). Using the OEM models (input [$\alpha$/Fe] = 0.1, [O/Fe] = 0.5), we estimate it to be a young star with mass = 0.90 $\pm$ 0.02 M$_{\odot}$, age = 5.68 $\pm$ 1.44 Gyr. The mean value of masses of OEM models ([O/Fe] = 0.5) inside the observational error box is larger than that of $\alpha$EM models ([O/Fe] = 0.1), leading to smaller modeling-determined age for this star.  Figure \ref{fig:grid-fit}(b) shows the tracks of a star with observed [$\alpha$/Fe] $\sim$ 0.2, [O/Fe] $\sim$ 0. We obtain a mass of 0.99 $\pm$ 0.01 M$_{\odot}$ and an age of 10.51 $\pm$ 0.60 Gyr for this star with $\alpha$EM models (input [$\alpha$/Fe] = 0.2, [O/Fe] = 0.2), and a mass of 0.98 $\pm$ 0.02 M$_{\odot}$ and an age of 11.34 $\pm$ 0.51 Gyr with OEM models (input [$\alpha$/Fe] = 0.2, [O/Fe] = 0). As seen, the OEM models with input [O/Fe] = 0 are generally hotter than the $\alpha$EM models ([O/Fe] = 0.2) at fixed mass and [Fe/H], leading to smaller modeling-determined mass and larger age for this star. 

\begin{figure*}[ht!]
\centering 
\includegraphics[scale=0.7]{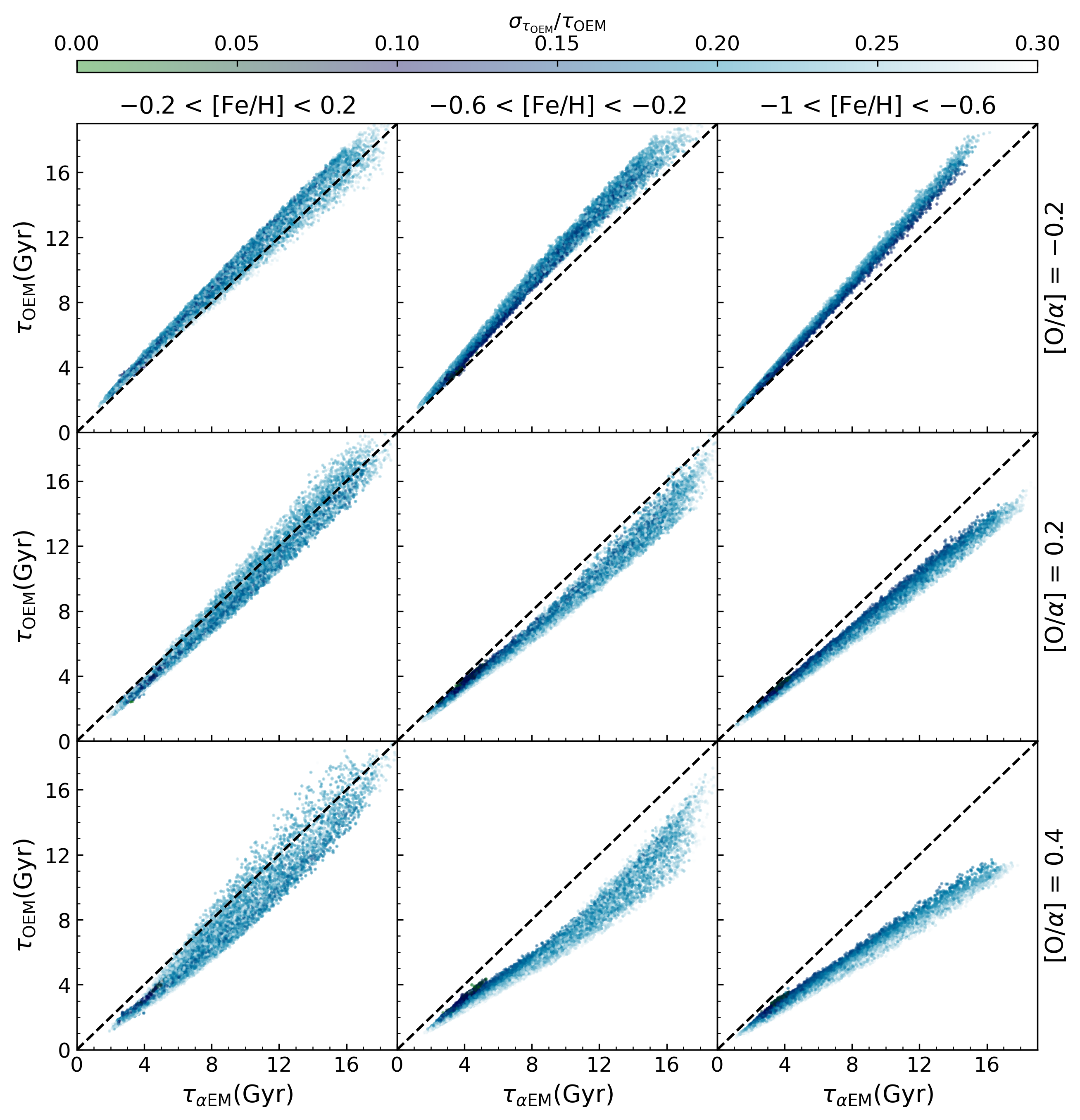} 
\caption{Comparison of the ages determined with $\alpha$EM and OEM models for mock stars, color-coded by
age uncertainty. These stars are divided by their [Fe/H] and
[O/$\alpha$] values. Black dash lines show the agonic line. 
\label{fig:o_0.2}}
\end{figure*}

\begin{figure*}[ht!]
\centering 
\includegraphics[scale=0.6]{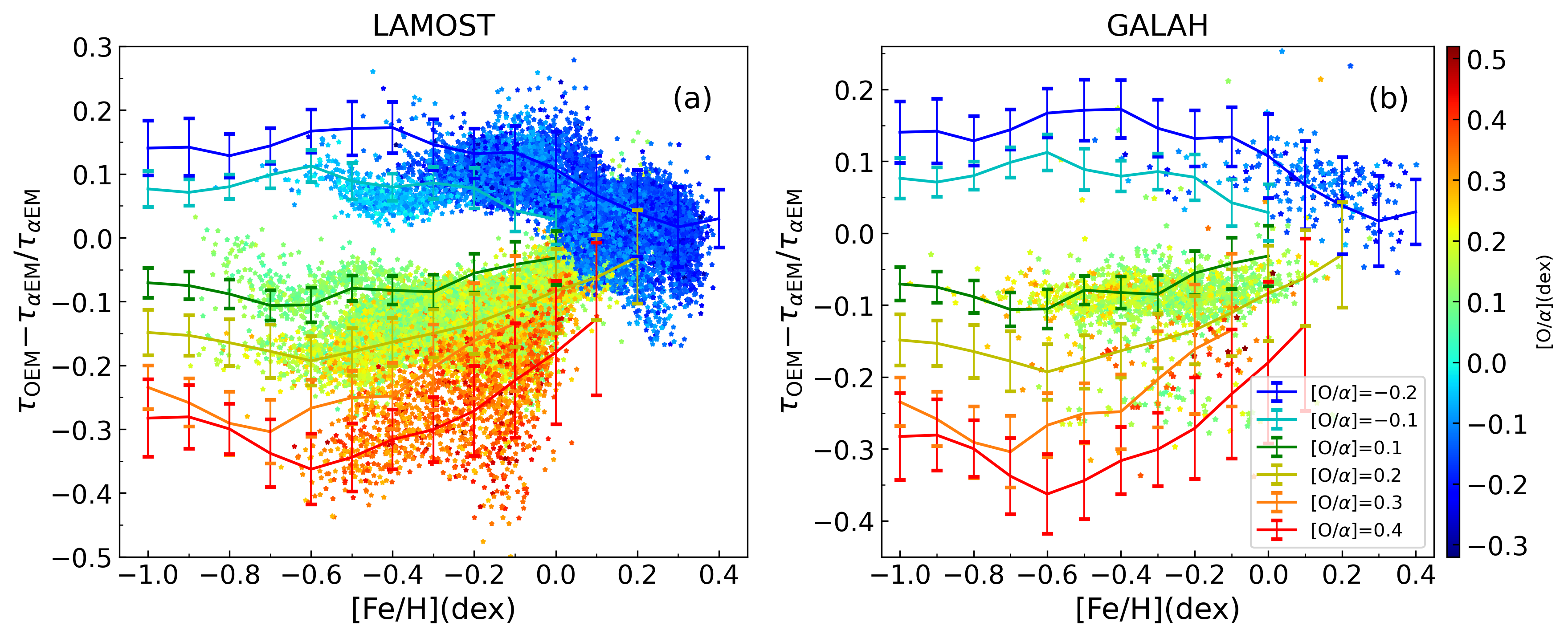} 
\caption{Distribution of fractional age difference as a function of [Fe/H] for sample stars from LAMOST (a) and GALAH (b), color-coded by [O/$\alpha$]. The overplotted solid lines represent the median and standard deviation of fractional age difference for fake stars in each [Fe/H] bin, with a bin size of 0.1 dex. \label{fig:change_all}}
\end{figure*}

\subsection{Fitting Method} \label{subsec:Parameter Estimation}

We constrain stellar masses and ages using five observed quantities, i.e., $T_{\rm eff}$, luminosity, [Fe/H], [$\alpha$/Fe], and [O/Fe]. Note that [O/Fe] is not used when estimating parameters with $\alpha$EM models.


We follow the fitting method raised by \cite{2010ApJ...710.1596B}. According to the Bayes theorem, we compare model predictions with their corresponding observational properties $D$ to calculate the overall probability of the model $M_i$ with posterior probability $I$,
\begin{equation}\label{e1}
p\left(M_{i}\mid D,I\right)=\frac{p\left(M_{i}\mid I\right) p\left(D\mid M_{i}, I\right)}{p(D\mid I)}
\end{equation}
where $p$($M_i$ $\mid$ $I$) represents the uniform prior probability for a specific model, and $p$(D $\mid$ $M_i$, $I$) is the likelihood function: 
\begin{equation}\label{e2}
\begin{aligned}
p\left(D\mid M_{i},I\right)=L(T_{eff},[Fe/H],lum)\\
=L_{T_{eff}}L_{[Fe/H]}L_{lum}
\end{aligned}
\end{equation}
The $p$($D$ $\mid$ $I$) in Equation \ref{e1} is a normalization factor for the specific model probability:
\begin{equation}\label{e4}
p(D \mid I)=\sum_{j=1}^{N_{m}} p\left(M_{j} \mid I\right) p\left(D \mid M_{j}, I\right)
\end{equation}
where $N_m$ is the total number of selected models. The uniform priors $p$($M_i$ $\mid$ $I$) can be canceled, giving the simplified Equation (1) as :
\begin{equation}\label{e5}
p\left(M_{i} \mid D, I\right)=\frac{p\left(D \mid M_{i}, I\right)}{\sum_{j=1}^{N_{m}} p\left(D \mid M_{j}, I\right)}.
\end{equation}
Then Equation \ref{e5} is the probability distribution for the selected models with the most probable fundamental parameters. 
As demonstrated in Figure \ref{fig:age-fit}, we fit a Gaussian function to the likelihood distribution of mass and age for each star.
The mean and standard deviation of the resulting Gaussian profile are then utilized as the median value and uncertainty of fundamental parameter (mass and age) for each star.
To find the stars that locate near the edge of the model grid, we consider a 3-sigma error box (i.e., three times the observational error, depicted as a blue square in Figure \ref{fig:grid-fit}) on the HR diagram and divide the error box into 100 bins. For a certain star, when there are more than 5 bins that do not contain any theoretical model (sampling rate $<$ 95\%), we flag the star with “edge effect”. 

To assess the accuracy of our models and investigate potential model dependency in age and mass determination, we present a comparison of results obtained from our $\alpha$EM models, OEM models, and the GALAH DR3 value-added catalog \citep[VAC,][]{2021MNRAS.506..150B}. Figure \ref{fig:age_a2} shows the comparison of age and mass estimations for $\sim$4,000 GALAH stars, with age uncertainty of less than 30\%, based on 
$\alpha$EM models, OEM models, and GALAH DR3 VAC \citep{2021MNRAS.506..150B}. The ages and masses of stars from GALAH DR3 VAC are calculated using the PARSEC (the PAdova and TRieste Stellar Evolution Code) release v1.2S + COLIBRI stellar isochrone \citep{2017ApJ...835...77M}, which adopt a solar-scaled metal mixture, i.e., input [$\alpha$/Fe] = 0. Figure \ref{fig:age_a2} illustrates that the one-to-one relation of the results is quite good for most stars. It is noteworthy that the adopted approach encompasses a flat prior on age with an age cap of 13.2 Gyr \citep{2018MNRAS.473.2004S}. Consequently, the ages of the majority of stars from GALAH DR3 VAC are found to be younger than 12 Gyr (with masses larger than 0.8 M$_{\odot}$), which results in a relatively large dispersion of age differences, amounting to 12.4\% for $\alpha$EM models and 13.0\% for OEM models. 
Significant systematic differences are apparent between the PARSEC and the $\alpha$EM models in Figure \ref{fig:age_a2}(a-b), with the former indicating 2.3\%  older age and 1.5\% smaller mass than the latter.
These discrepancies could be attributed to differences in the input physics employed by the two models, such as the input [$\alpha$/Fe] value, helium abundance, and mixing-length parameter.
In Figure \ref{fig:age_a2}(c-d), the PARSEC yields 5.5\% older age and 1.9\% smaller mass than the OEM models.
Compared with the $\alpha$EM models, the OEM models demonstrate more pronounced systemic differences from PAESEC. These distinctions primarily arise from the  consideration of O-enhancement in OEM models, leading to younger ages and higher masses.
In addition, a comparison of results obtained from our $\alpha$EM models and the Yonsi–Yale \citep[YY,][]{2008IAUS..252..413Y} stellar isochrones have been shown in Figure \ref{fig:age_a1} in Appendix.

\section{Results} \label{sec:result}

This work aims to determine the ages of dwarfs considering oxygen abundance and study the chemical and kinematic properties of high-$\alpha$ and low-$\alpha$ populations in the Galactic disk. We give the masses and ages of 149,906 LAMOST dwarfs and 15,591 GALAH dwarfs with $\alpha$EM models and OEM models. We remove $\sim$30\% stars with sampling rate $<$ 95\%, located near the edge of the model grid. In addition, we remove $\sim$3\% stars whose inferred ages are 2-sigma\footnote{For a certain star, age $-$ 2*age$\_$uncertainty $>$ 13.8 Gyr.} larger than the universe age \citep[13.8 Gyr,][]{2016A&A...594A..13P} due to their significant model systematic bias. Finally, we remove $\sim$35\% stars that have relative age uncertainty larger than 30 percent. After these cuts, we obtain the ages of 67,503 dwarfs from LAMOST with a median age uncertainty of $\sim$16\%, and 4,006 dwarfs from GALAH with a median age uncertainty of $\sim$18\%.
The age estimation of dwarf stars is inherently accompanied by considerable uncertainty, which can reach up to 30\% within our sample. Furthermore, uncertainties (especially the systematic error) in atmosphere parameters can introduce biases in the age estimation. Consequently, a minority of stars in our sample exhibits ages that exceed the age of the universe. This occurrence is not uncommon, as even samples of subgiants with more precise age determinations have encountered analogous occurrences \citep{2022Natur.603..599X}.

\subsection{Oxygen Effect on Age Determinations} \label{sec:oxygen_effect}

\subsubsection{Mock Data Test} \label{sec:mock_data}


Most of the stars in both the LAMOST and GALAH samples are distributed in a relatively narrow range of [Fe/H] ($-$0.5 dex - +0.5 dex).
To systematically investigate the effect of O-enhancement on age determinations in a wide range of $T_{\rm eff}$ and [Fe/H], we apply a mock data test based on our grid of stellar models. For each set of stellar mode grids with fixed [Fe/H], [$\alpha$/Fe], and [O/Fe] values, we draw random samples from the distributions of stellar evolution tracks in the H-R diagram. 
We adopt 0.05, 30 K as the observational errors for [Fe/H] and $T_{\rm eff}$, and fractional error of 2\% for luminosity. Finally, We generate mock data of 0.15 million stars with age uncertainty of less than 30 percent. 

Figure \ref{fig:mock_data_test}(a) shows the distribution of mock stars on the HR diagram. Figure \ref{fig:mock_data_test}(b-c) presents a comparison between mock data and observational data for $T_{\rm eff}$ and [Fe/H] distributions. Comparing mock data with LAMOST or GALAH dwarfs, mock stars cover wider ranges of $T_{\rm eff}$ (5000 K - 7000 K), and [Fe/H] ($-$1.0 dex - +0.4 dex). Therefore, the mock data is useful for statistical studies of oxygen effect on age determinations.

Figure \ref{fig:o_0.2} shows a comparison between ages determined with $\alpha$EM models ($\tau_{\alpha\rm EM}$) and OEM models ($\tau_{\rm OEM}$). The mock stars are grouped by their [Fe/H] and [O/$\alpha$] values. The stars with [O/$\alpha$] $>$ 0 are hereafter referred to as high-O stars and the stars with [O/$\alpha$] $<$ 0 as low-O stars. Generally, high-O stars have younger ages based on OEM models, while low-O stars become older. The effect of oxygen enhancement on age determination is relatively significant for stars with [Fe/H] $<$ $-$0.2. At [O/$\alpha$] = $-$0.2, the mean fractional age difference ( ($\tau_{\rm OEM}$ $-$ $\tau_{\alpha\rm EM}$)/$\tau_{\alpha\rm EM}$ ) is 10.5\% for metal-rich stars ($-$0.2 $<$ [Fe/H] $<$ 0.2), and 15.5\% for relatively metal-poor stars ($-$1 $<$ [Fe/H] $<$ $-$0.2). The mean fractional age difference at [O/$\alpha$] = 0.2 is $-$9.2\% for metal-rich stars, and $-$16.5\% for relatively metal-poor stars.
The largest fractional age difference comes from high-O stars with [O/$\alpha$] = 0.4, which have a mean fractional age difference of $-$20.2\% at $-$0.2 $<$ [Fe/H] $<$ 0.2, and $-$30.6\% at $-$1 $<$ [Fe/H] $<$ $-$0.2. 
We find clear age offsets that correlate to the [Fe/H] and [O/$\alpha$] values. Increasing 0.2 dex in [O/$\alpha$] will reduce the age estimates of metal-rich stars by $\sim$10\%, and metal-poor stars by $\sim$15\%.
The mock data provide us with more sufficient stars at the metal-poor edge than observational data to present clearly age differences at different [O/$\alpha$] and [Fe/H] values.

\subsubsection{Observational Data} \label{sec:observational_data}

Figure \ref{fig:change_all} presents the fractional age differences between $\alpha$EM and OEM models for observational (LAMOST and GALAH) and mock data. The overall average age offset (absolute value of age difference) of stars from LAMOST and GALAH is 8.9\% and 8.6\%, respectively. Of the low-O stars with [Fe/H] $<$ 0.1 dex and [O/$\alpha$] $\sim$ $-$0.2 dex, many have fractional age differences of $\geq$ 10\%, and even reach up to 27\%. The mean fractional age difference of high-O stars with [O/$\alpha$] $\sim$0.4 dex is $\sim$ $-$25\%. The age offsets are relatively significant for metal-poor stars. The largest age differences are $-$33\% to $-$42\% for stars with [Fe/H] $\lesssim$ $-$0.6 dex and [O/$\alpha$] $\sim$0.4 dex. For mock data, we note the trend of age offsets versus [Fe/H] is consistent with that of observational data. The age offsets of both samples increase significantly with decreasing metallicity at [Fe/H] $\gtrsim$ $-$0.6. Interestingly, there is a slight increase in age offsets with decreasing metallicity at [Fe/H] $<$ $-$0.6. 
This trend of age offsets is consistent with the change of $T_{\rm eff}$ difference as a function of [Fe/H] (shown in Figure \ref{fig:grid}), as discussed in Section \ref{subsec:grid}.

\begin{figure*}[ht!]
\includegraphics[scale=0.58]{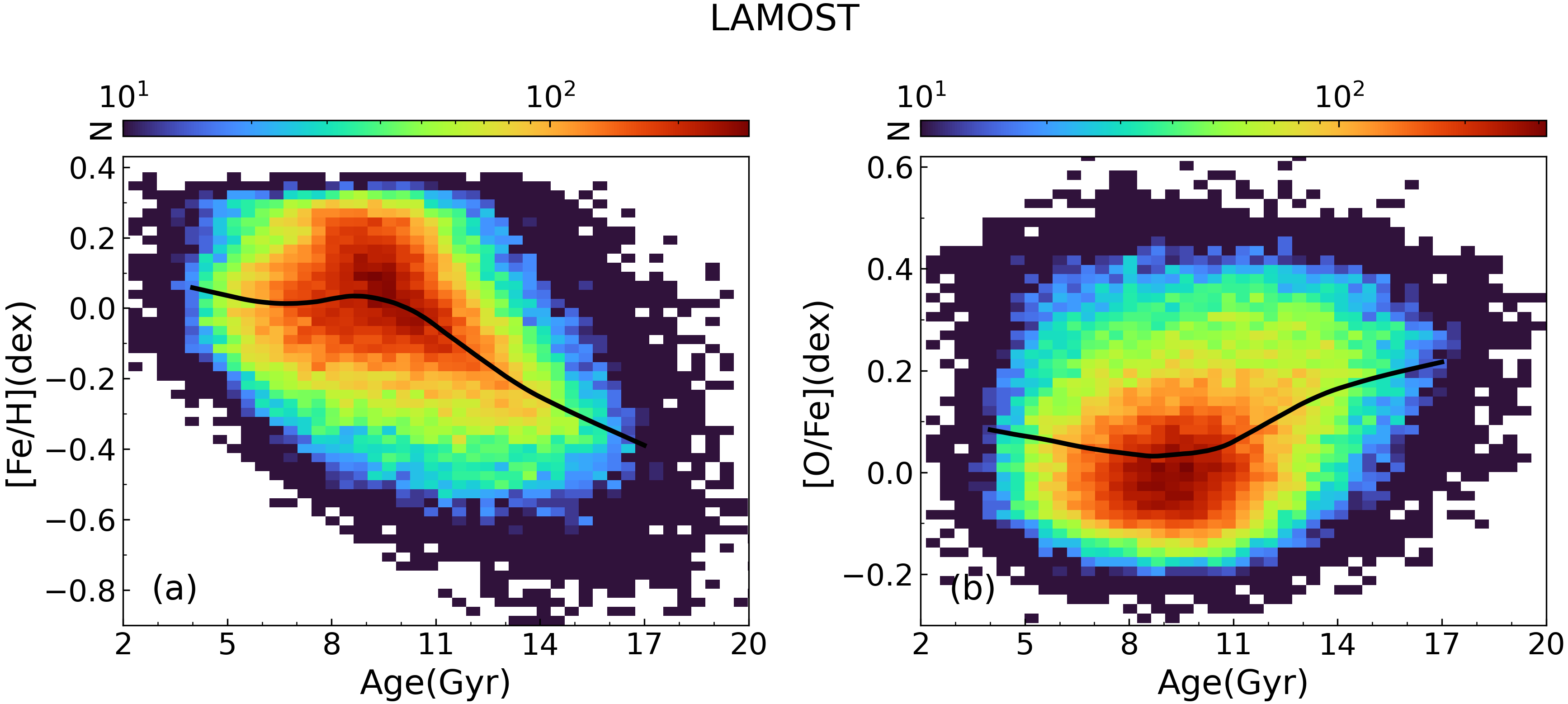}
\caption{Color-coded stellar number density distribution of the sample stars from LAMOST in the age-[Fe/H] (a) and age-[O/Fe] (b) plane. The black solid lines represent the
fitting for age-abundance relations by local nonparametric regression. \label{fig:age_bin2}}
\end{figure*}

\begin{figure*}[ht!]
\includegraphics[scale=0.58]{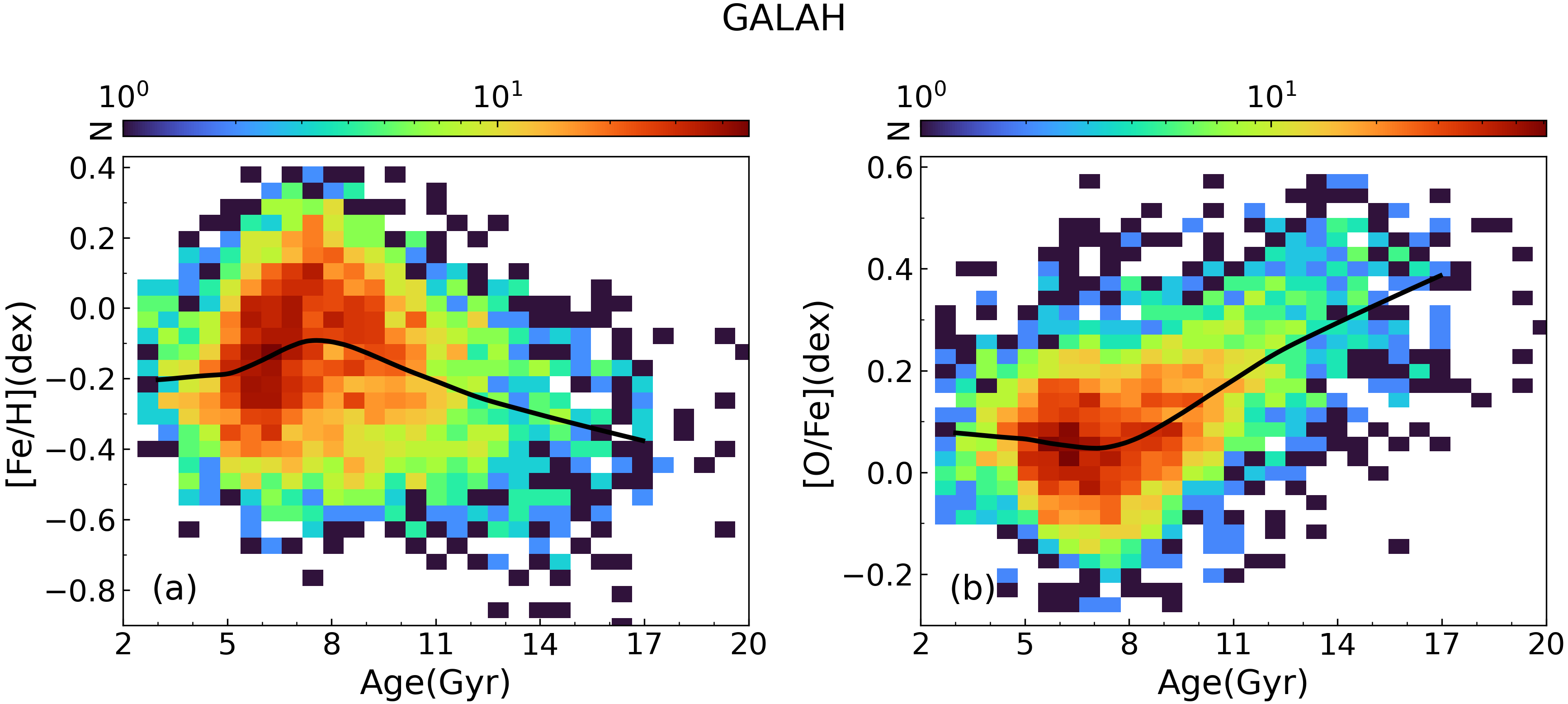}
\caption{Color-coded stellar number density distribution of the sample stars from GALAH in the age-[Fe/H] (a) and age-[O/Fe] (b) plane. The symbols are the same as those defined in Figure \ref{fig:age_bin2}.  \label{fig:age_bin2_GALAH}}
\end{figure*}

\subsection{Age-Abundance Relations} \label{sec:discussion}

To trace the chemical evolution history of the Galactic disk, we hereby present the age-abundance relations of the LAMOST sample (consisting of 67,511 stars) and the GALAH sample (consisting of 4,006 stars) using the ages from OEM models. For each sample, we employ local nonparametric regression fitting (LOESS model) to characterize the trends in these relations with enhanced clarity.

Figure \ref{fig:age_bin2} illustrates the results for the LAMOST sample. In Figure \ref{fig:age_bin2}(a), a gradual decline in [Fe/H] is observed across the age range of $\sim$9 Gyr to $\sim$6.5 Gyr. This trend shows similarities to the metal-rich branch observed in young stars (age $<$ 8 Gyr) as found by \citet{2022Natur.603..599X}, where the metallicity range of their metal-rich branch stars spans approximately $-$0.2 to +0.4. Notably, \citet{2022MNRAS.510.4669S} also identifies a trend comparable to our findings, whereby their sample exhibits a [Fe/H] value of 0.4 at 8 Gyr, diminishing to around $-$0.2 at 6 Gyr. The "two-infall" chemical evolution model \citep{1997ApJ...477..765C,2017MNRAS.472.3637G} predicts a process involving the infall of metal-poor gas commencing roughly 9.4 Gyr ago \citep{2019A&A...623A..60S,2020A&A...635A..58S}. The observed trend of decreasing metallicity from 9 Gyr to 6.5 Gyr in our results may be related to this infalling metal-poor gas. Intriguingly, this "two-infall" model not only anticipates a decline in metallicity but also predicts an increase in the oxygen abundance,which is consistent with the observed trend illustrated in Figure \ref{fig:age_bin2}(b). In Figure \ref{fig:age_bin2}(b), the sample stars from LAMOST exhibit an increase in [O/Fe] as the age decreases from 9 Gyr to 4 Gyr, indicating a slight enrichment of oxygen in the younger stellar population.

Figure \ref{fig:age_bin2_GALAH} presents the results for the GALAH sample. It is noteworthy that the GALAH stars display a decrease in [Fe/H] from $\sim$7.5 Gyr to 5 Gyr. Furthermore, the [O/Fe] of the GALAH stars exhibit a slight decrease with age ranging from $\sim$7.5 Gyr to 3 Gyr. The GALAH sample exhibits age-[Fe/H] and age-[O/Fe] trends similar to those observed in LAMOST; however, an overall slight temporal discrepancy can be observed. This incongruity may be ascribed to dissimilarities in sample composition or systematic differences in atmospheric parameters between the two survey datasets. The GALAH sample, on the whole, exhibits higher temperatures compared to LAMOST sample (5000 - 5700 K), indicating a relatively younger population. Furthermore, the determinations of [Fe/H] and [O/Fe] from GALAH are based on a non-LTE method \citep{2020A&A...642A..62A}, which can also impact the observed trends.

In conclusion, the analysis of the LAMOST and GALAH samples reveals a decreasing trend of [Fe / H] with an age ranging from 7.5--9 Gyr to 5--6.5 Gyr, and a notable upward trend in [O/Fe] as the age decreases from 7.5--9 Gyr to 3--4 Gyr. This result agree with the prediction of the "two-infall" scenario and suggest that a metal-poor and O-rich gas gradually dominates the star formation from 7.5--9 Gyr ago. As discussed in Section \ref{sec:intro}, oxygen has a unique origin, primarily produced by CCSNe \citep{2021AJ....161....9F}. Therefore, the observed age-[O/Fe] trend plays a distinct role in characterizing the chemical evolution history of the Milky Way and constraining chemical evolution models. 
Neglecting to account for the independent enhancement of oxygen abundance in age determination would result in significant age biases, as discussed in Section \ref{sec:oxygen_effect}. Such biases would obscure the age-[O/Fe] relation, as depicted in Figure \ref{fig:age_bin2_aem} in the appendix, where the rising trend of [O/Fe] with decreasing age remains imperceptible at age $<$ 9 Gyr. Therefore, we suggest that considering the oxygen abundance independently in stellar models is crucial. This would aid in accurately characterizing the age-[O/Fe] relation and provide better constraints for Galactic chemical evolution models.

\section{CONCLUSIONS}
\label{sec:conclusion}

To determine the ages of dwarfs considering observed oxygen abundance, we construct a grid of stellar models which take into account oxygen abundance as an independent model input. We generate mock data with 0.15 million mock stars to systematically study the effect of oxygen abundance on age determination. Based on the $\alpha$-enhanced models and O-enhanced models, we obtain the masses and ages of 67,503 stars from LAMOST and 4,006 stars from GALAH and analyze the chemical and kinematic properties of these stars combined with ages from O-enhanced models.

Our main conclusions are summarized as follows:

(\romannumeral1) The ages of high-O stars based on O-enhanced models are smaller compared with those determined with $\alpha$-enhanced models, while low-O stars become older. We find clear age offsets that correlate to the [Fe/H] and [O/$\alpha$] values. Varying 0.2 dex in [O/$\alpha$] will alter the age estimates of metal-rich ($-$0.2 $<$ [Fe/H] $<$ 0.2) stars by $\sim$10\%, and relatively metal-poor ($-$0.2 $<$ [Fe/H] $<$ 0.2) stars by $\sim$15\%.

(\romannumeral2) The overall average age offset (absolute value of age difference) between $\alpha$-enhanced models and O-enhanced models is 
8.9\% for LAMOST stars, and 8.6\% for GALAH stars. Of the low-O stars with [Fe/H] $<$ 0.1 dex and [O/$\alpha$] $\sim$ $-$0.2 dex, many have fractional age differences of $\geq$ 10\%, and even reach up to 27\%. The mean fractional age difference of high-O stars with [O/$\alpha$] $\sim$0.4 dex is $\sim$ $-$25\%, and reach up to $-$33\% to $-$42\% at [Fe/H] $\lesssim$ $-$0.6 dex.

(\romannumeral3) Based on LAMOST and GALAH samples, we observe a decreasing trend of [Fe/H] with age from 7.5--9 Gyr to 5--6.5 Gyr. Furthermore, The [O/Fe] of both sample stars increases with decreasing age from 7.5--9 Gyr to 3--4 Gyr, which indicates that the younger population of these stars is more O-rich. Our results agree with the prediction of the "two-infall" scenario and suggest that a metal-poor and O-rich gas gradually dominates the star formation from 7.5--9 Gyr ago.


We thank the anonymous referee for valuable comments and suggestions that have significantly improved the presentation of the manuscript. This work is based on data acquired through the Guoshoujing Telescope. Guoshoujing Telescope (the Large Sky Area Multi-Object Fiber Spectroscopic Telescope; LAMOST) is a National Major Scientific Project built by the Chinese Academy of Sciences. Funding for the project has been provided by the National Development and Reform Commission. LAMOST is operated and managed by the National Astronomical Observatories, Chinese Academy of Sciences.
This work used the data from the GALAH survey, which is based on observations made at the Anglo Australian Telescope, under programs A/2013B/13, A/2014A/25, A/2015A/19, A/2017A/18, and 2020B/23.
This work has made use of data from the European Space Agency (ESA) mission Gaia (\url{https://www.cosmos.esa.int/gaia}), processed by the Gaia Data Processing and Analysis Consortium (DPAC, \url{https://www.cosmos.esa.int/web/gaia/dpac/consortium}). Funding for the DPAC has been provided by national institutions, in particular the institutions participating in the Gaia Multilateral Agreement.
This work is supported by National Key R$\&$D Program of China No. 2019YFA0405503,  the Joint Research Fund in Astronomy (U2031203,) under cooperative agreement between the National Natural Science Foundation of China (NSFC) and Chinese Academy of Sciences (CAS), and NSFC grants (12090040, 12090042). This work is partially supported by the CSST project, and the Scholar Program of Beijing Academy of Science and Technology (DZ:BS202002). This paper has received funding from the European Research Council (ERC) under the European Union’s Horizon 2020 research and innovation programme (CartographY GA. 804752).

\appendix

\begin{figure*}[ht]
\includegraphics[scale=0.55]{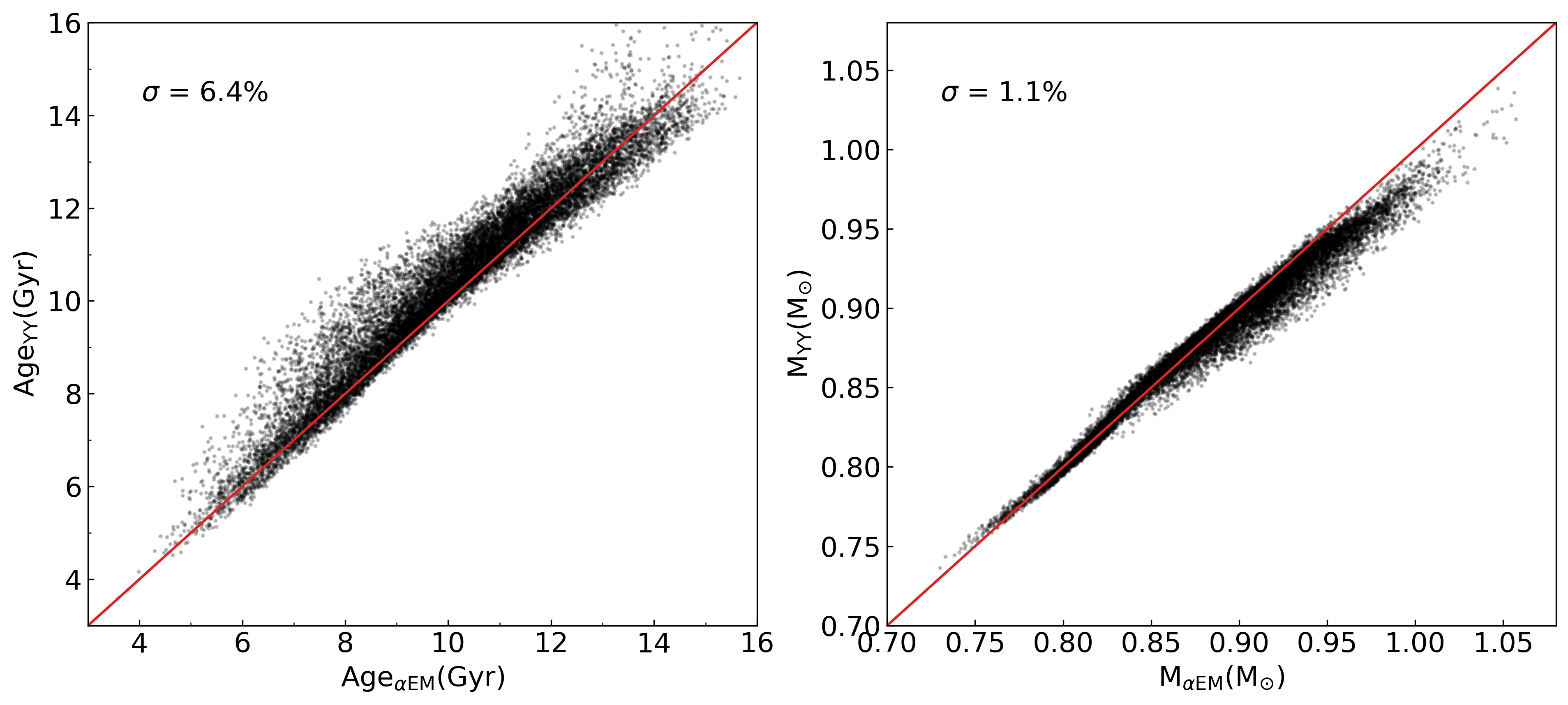}
\caption{Comparison of age and mass estimates for $\sim$15,000 LAMOST stars using our $\alpha$EM models and the Yonsi–Yale \citep[YY,][]{2008IAUS..252..413Y} stellar isochrones with input [$\alpha$/Fe] = 0.1 dex. The red line represents the 1:1 line. Dispersion of the relative age and mass difference is marked in the figure. \label{fig:age_a1}}
\end{figure*}

Figure \ref{fig:age_a1} depicts the age and mass determinations for $\sim$15,000 LAMOST stars (with [$\alpha$/Fe] $\sim$ 0.1) and reveals a satisfactory correspondence between the $\alpha$EM models and the YY isochrones \citep{2008IAUS..252..413Y}, as the dispersion of the relative age and mass differences are only 6.4\% and 1.1\% between these two models. However, slight systematic differences are visible among this result, as the YY yields 3.6\% older age and $-$0.4\% smaller mass than the $\alpha$EM models. 

\begin{figure*}[ht!]
\includegraphics[scale=0.58]{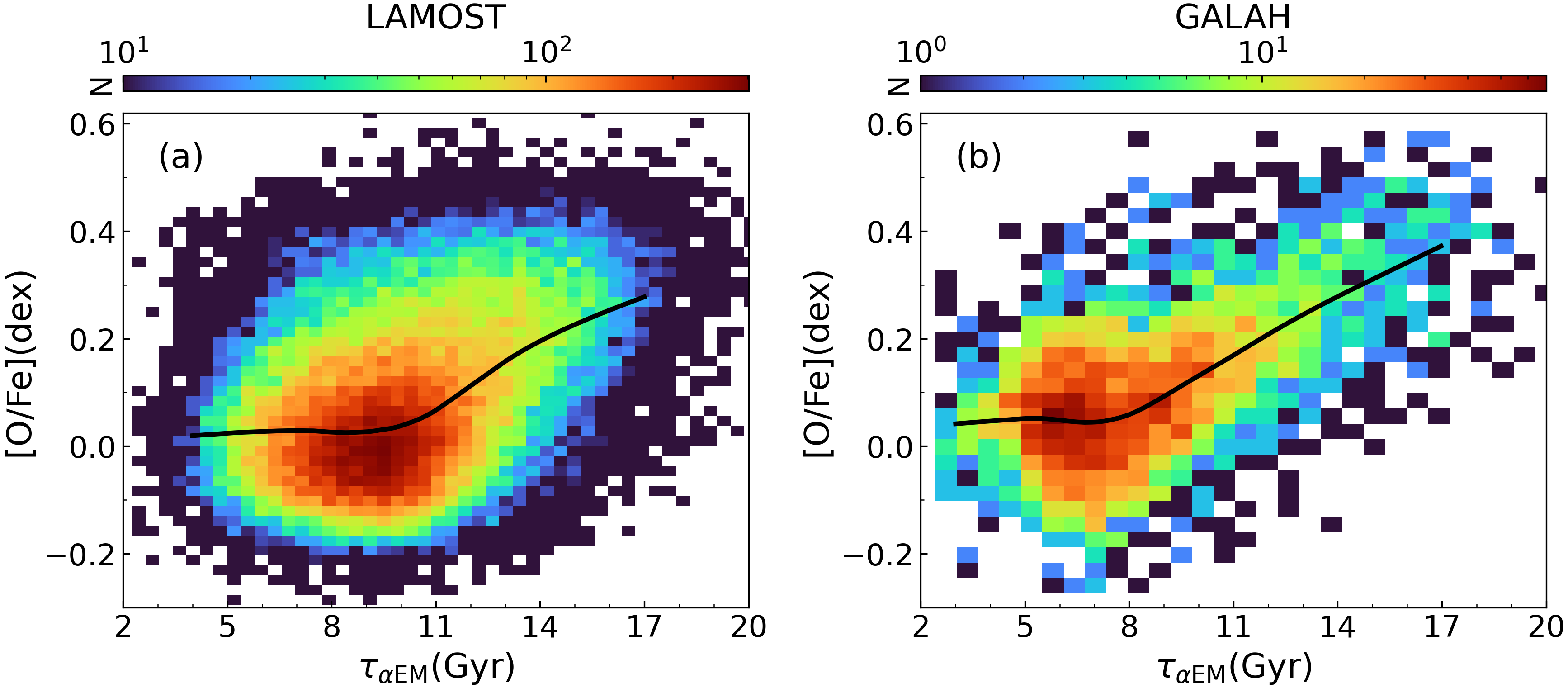}
\caption{Color-coded stellar number density distribution of the sample stars from LAMOST (a) and GALAH (b) in the age-[O/Fe] plane, based on the ages from $\alpha$EM models. The black solid lines represent the fitting for age-abundance relations by local nonparametric regression. \label{fig:age_bin2_aem}}
\end{figure*}


\bibliography{ref}{}
\bibliographystyle{aasjournal}



\end{document}